
\input phyzzx.tex
\pubnum{ROM2F-95-12}
\unnumberedchapters
\def\G{\Gamma}

\def\a{\alpha}
\def\b{\beta}

\def\eps{\epsilon}
\def\vareps{\varepsilon}

\def\l{\lambda}
\def\cd{{\cal D}}
\def\ct{{\cal T}}
\def\m{\mu}
\def\n{\nu}

\def\s{\sigma}
\def\t{\tau}
\def\z{\zeta}
\def\w{\omega}
\def\mn{\mu\nu}

\def\xbar{\bar x}

\def\ddr{{\partial \over \partial r}}

\def\adot{\dot\alpha}

\def\mezzo{{1 \over 2}}
\def\unoar{1-\bigl({a \over r}\bigr)^4}

\def\ap{\alpha^\prime~}
\def\ALE{Asymptotically~Locally~Euclidean~}

\def\EH{Eguchi-Hanson~}

\def\YM{Yang-Mills~}

\def\KN{Kronheimer-Nakajima~}
\def\Kro{Kronheimer~}
\def\Nak{Nakajima~}
\def\GH{Gibbons-Hawking~}

\def\tb{tautological~bundle~}
\def\ms{moduli~space~}

\def\zms{zero-modes~}
\def\th{triholomorphic~}

\def\hkq{hyperk\"ahler~quotient~}
\def\hk{hyperk\"ahler~}
\def\hkm{hyperk\"ahler~manifold~}

\def\sd{self-dual~}

\def\stand{standard~embedding~}
\def\min{minimal~instanton~}
\def\mc{multicenter~}
\def\mom{moment~map~}
\def\unita{{1 \kern-.30em 1}}
\def\fey{{\big / \kern-.80em D}}
\def\complex{{\kern .1em {\raise .47ex \hbox {$\scriptscriptstyle |$}}
\kern -.4em {\rm C}}}
\def\zet{{Z \kern-.45em Z}}
\def\real{{\vrule height 1.6ex width 0.05em depth 0ex \kern -0.06em {\rm R}}}
\def\rational{{\kern .1em {\raise .47ex \hbox{$\scripscriptstyle |$}}
\kern -.35em {\rm Q}}}
\titlepage
\title{Instanton Effects in Supersymmetric Yang-Mills Theories
on ALE Gravitational Backgrounds}\foot{Work partially supported by E.C.
Grant CHRX-CT93-0340.}
\author{Massimo~Bianchi, Francesco~Fucito, Giancarlo~Rossi}
\address{Dipartimento di Fisica, Universit\`a di Roma II ``Tor
Vergata'' and I.N.F.N. Sezione di Roma II ``Tor Vergata'',00133 Roma, Italy}
\andauthor{Maurizio~Martellini}
\address{Dipartimento di Fisica, Universit\`a di Milano, 20133
Milano, Italy and Sezione I.N.F.N. dell'Universit\`a di Pavia,
27100 Pavia, Italy}

\abstract
{In this letter we report on the computation of
instanton-dominated correlation functions in supersymmetric \YM theories on
\ALE spaces.
Following the approach of Kronheimer and Nakajima, we explicitly
construct the self-dual connection on \ALE spaces necessary to perform such
computations. We restrict our attention to the simplest case of an
$SU(2)$ connection with lowest Chern class on the \EH gravitational
background.}
\endpage
\pagenumber=1

\leftline{\bf Introduction}
\REF\susy{For a review see D.~Amati, K.~Konishi, Y.~Meurice, G.C.~Rossi and
G.~Veneziano, Phys. Rep. {\bf 162} (1988) 169 and references therein.}
\REF\sw{N.~Seiberg and E.~Witten, Nucl. Phys. {\bf B426} (1994) 19;
ibid. {\bf B431} (1994) 484.}
\REF\ggpz{L.~Girardello, A.~Giveon, M.~Porrati and A.~Zaffaroni,
{\it S-Duality in $N=4$ Yang-Mills Theories with General Gauge Group},
preprint hep-th/9502057;
{\it Non-Abelian Strong-Weak Coupling Duality in (String-Derived) $N=4$
Supersymmetric Yang-Mills Theories}, preprint hep-th/9406128.}
\REF\wym{E.~Witten, {\it Supersymmetric Yang-Mills Theory on a Four-Manifold},
preprint IASSNS-HEP-94/5.}
\REF\vw{C.~Vafa, E.~Witten, {\it A Strong Coupling Test of S-Duality}, preprint
hep-th/9408074.}
Understanding non-perturbative phenomena is a key issue in any field
theory. These effects are supposed to play a fundamental role
in the explanation of confinement as well as dynamical supersymmetry (SUSY)
breaking and many interesting results that go beyond perturbation theory
have been recently obtained for supersymmetric
\YM theory (SYM from now on) on flat [\susy,\sw,\ggpz] and curved
manifolds [\wym,\vw].

\REF\wtop{E.~Witten, Comm. Math. Phys. {\bf 117} (1988) 353.}
\REF\dk{S.K.~Donaldson and P.B.~Kronheimer, {\it The Geometry of
Four-Manifolds}, Oxford University Press 1990.}
Constant values for instanton dominated correlation functions
may be related to topological invariants of the \ms of YM connections
on the base manifold [\wtop,\dk]. When clustering applies, they give
rise to the formation of chiral condensates.

The local extension of the results obtained for globally SYM theories
presents formidable difficulties, the main one being the non renormalizability
of the resulting quantum (super-)gravity. A way to circumvent this problem
is to have the theory embedded in a suitable string theory
which will act as a regulator. One thus seems to be lead to
consider only the four dimensional effective field theories which emerge
as low-energy limits of consistent superstring compactifications.

\REF\chs{C.~Callan, J.~Harvey and A.~Strominger,
Nucl. Phys. {\bf B359} (1991) 611; ibid. {\bf B367} (1991) 60.}
\REF\khu{R.R.~Khuri, Nucl. Phys. {\bf B387} (1992) 315.}
\REF\bfmr{M.~Bianchi, F.~Fucito, M.~Martellini, G.C.~Rossi,
Nucl. Phys. {\bf B440} (1995) 195.}
\REF\kal{E.~Bergshoeff, R.~Kallosh and T.~Ortin, Phys. Rev. {\bf D51} (1995)
3009; R.~Kallosh, D.~Kastor, T.~Ortin, T.~Torma, Phys. Rev. {\bf D50} (1994)
6374.}
\REF\dkl{For a review see M.G.~Duff, R.R.~Khuri, J.X.~Lu, {\it String
Solitons},
preprint hep-th/9412184.}
\REF\rey{S.-J.~Rey, Phys. Rev. {\bf D43} (1991) 526.}
\REF\afhyp{D.~Anselmi and P.~Fre', Nucl. Phys. {\bf B404} (1993) 288;
ibid. {\bf B416} (1994) 255; Phys. Lett. {\bf 347B} (1995) 247.}
\REF\bdffv{M.~Billo', R.~D'Auria, S.~Ferrara, P.~Fre', P.~Soriani and
A.~Van Proeyen, {\it R-Symmetry and the Topological Twist of $N=2$ Effective
Supergravities of Heterotic Strings}, preprint hep-th/9505123.}
\REF\wmon{E.~Witten, Math. Res. Lett. {\bf 1} (1994) 769.}
Euclidean supersymmetric solutions of the heterotic (or Type I) string
equations of motion can be found
setting to zero the fermionic fields together with their
SUSY variations.
For these solutions not to get corrections in
(the $\s$-model coupling constant)
$\ap$, one is lead to impose the ``standard-embedding'' of
the generalized
spin connection into the gauge group [\chs,\khu,\bfmr,\kal,\dkl].
Leaving aside solutions with non-trivial configurations of the scalar fields
[\rey,\afhyp,\bdffv],
which may be related to monopole solutions [\wmon], one has
the option of either taking a constant dilaton background [\bfmr]
or a non-trivial axionic instanton [\rey,\chs,\khu].
\REF\hita{N.~Hitchin, Math. Proc. Camb. Phil.
Soc. (1979) {\bf 85}, 465.}
\REF\kro{P.B.~Kronheimer, J. Diff. Geom. {\bf 29} (1989) 665.}
\REF\adhm{M.~Atiyah, V.~Drinfeld, N.~Hitchin and Y.~Manin, Phys. Lett. {\bf
65A}
(1978) 185.}
\REF\kn{P.B.~Kronheimer and H.~Nakajima, Math. Ann. {\bf 288} (1990) 263.}
\REF\abfgz{D.~Anselmi, M.~Billo', P.~Fre', L.~Girardello, A.~Zaffaroni,
Int. Jou. Mod. Phys. {\bf A9} (1994) 3007.}

In the following we will mainly concentrate on the former choice
which leads to self-dual gauge connections on manifolds with self-dual
curvature. We further restrict our investigation to self-dual ALE
instantons. They have been completely classified
by Hitchin and Kronheimer [\hita,\kro] and, as shown by \Kro and \Nak [\kn],
admit a generalized ADHM procedure [\adhm] which can be exploited to
completely solve the problem of constructing the
most general self-dual gauge connection on an ALE manifold.
String propagation on ALE manifolds has been considered in [\bfmr,\abfgz].
It should be remembered that the metrics on these manifolds represent true
minima of the gravitational part of the locally supersymmetric action
which is obtained as a low-energy limit of the heterotic (or Type I) string.
ALE instantons can thus be taken as
good starting points for interesting string-inspired calculations
(\eg~for the study of dynamical supersymmetry breaking).

\REF\bfmrb{M.~Bianchi, F.~Fucito, M.~Martellini, G.C.~Rossi,
in preparation.}
The purpose of this letter is to briefly illustrate
the computation of instanton-dominated correlation
functions of SYM theories on ALE manifolds.
For definitiveness we will consider
the simplest case: an $SU(2)$ YM instanton with the lowest possible Chern class
(\ie~$c_2=\mezzo$) on the \EH gravitational background
(from now on we will refer to this case as the ``minimal instanton'').
A general description of the techniques
underlying the \KN (KN) construction as well as the much more complicated
computations involving $SU(2)$ istantons with higher values of the second Chern
class will be presented elsewhere [\bfmrb].
We would like to stress here that the \min is a solution
of the heterotic string equations of motion
only to lowest order in $\ap$. The constant dilaton background must be
corrected order by order in $\ap$.
On the contrary the solution obtained through the \stand [\bfmr],
which corresponds to $c_2={3 \over 2}$, is not expected to receive perturbative
corrections in $\ap$ and will be discussed in [\bfmrb].

\REF\nak{H.~Nakajima, Invent. Math. {\bf 102} (1990) 267.}
The four-dimensional
``framed" moduli space\foot{The ``framed" moduli space is defined as
the space of self-dual connections modulo the group of
homotopically trivial gauge transformations with local support.}
of the \min turns out to be equivalent, as a \hkm, to the base
manifold itself [\kn,\nak]. The correlation functions
which saturate the chiral selection rule depend on the number of
supersymmetries. For $N=1$ it is the vacuum expectation value
({\it vev}) of the
one-point gaugino bilinear ${\rm Tr}(\l\l(x))$ which is relevant.
For $N=2$ one has to consider the one-point composite
field ${\rm Tr}(\phi^2(x))$.
Finally, for $N=4$, it is simply the {\it vev} of the identity
operator which receives contribution from the minimal instanton.
\REF\eh{T.~Eguchi and A.J.~Hanson, Phys. Lett. {\bf 74B} (1978) 249;
Ann. Phys. {\bf 120} (1979) 82.}

The prototype of ALE metrics is the \EH (EH) gravitational background [\eh]
$$
ds^2= g^{EH}_{\mn} dx^\mu dx^\nu=
({r\over u})^2 dr^2+r^2 (\sigma_x^2+\sigma^2_y)+ u^2 \sigma_z^2
\eqn\ehmetric
$$
where $\s_x$, $\s_y$ and $\s_z$ are the left-invariant forms on $SU(2)$.
The bolt singularity at $r=a$, is removed by changing
the radial variable to
$u=r \sqrt{\unoar}$ and identifying antipodal points. The EH background
has an $S^3/\zet_2$ boundary and admits
an $SU(2)_R \otimes U(1)_L$ isometry group. It is well known
that the EH metric is a solution of the euclidean Einstein equations
with Euler characteristic, $\chi$,
equal to 2 and Hirzebruch signature, $\tau= b^+_2-b^-_2$, equal to 1.
Coupling this solution to gauge fields via the ``standard embedding",
allows to promote it to a fullfledged solution of the heterotic
string equations of motion [\bfmr].

We recall that, as there are no (normalizable) anti-self-dual two-forms
on the EH background,
there exists exactly one self-dual two-form given by [\eh]
$$
F = F_{\mn} dx^\mu \wedge dx^\nu=
{2a^2\over r^4} (r dr \wedge \s_z + r^2 \s_x \wedge \s_y )
\eqn\maxwell
$$
which may be thought as the field strength of the ``monopole potential''
$$
A = A_\m dx^\m = {a^2\over r^2} \s_z
\eqn\ehmonopole
$$

\REF\gh{G.~Gibbons, S.~Hawking, Phys. Lett. {\bf 78B} (1978) 430.}
Gibbons and Hawking [\gh] generalized the metric \ehmetric~ to a class
of \mc metrics with increasing Euler characteristic and Hirzebruch signature.
The EH case corresponds to a ``two-center'' metric, the ``one-center'' case
being diffeomorphic to flat Euclidean space. Eliminating the apparent
singularities in the $n$-center metrics at the location of the centers,
requires
an identification of points under a discrete group $\zet_n$ so that the
boundary
of the manifold turns out to be $S^3/\zet_n$ and the resulting asymptotic
metric is not globally Euclidean but only Asymptotically Locally Euclidean,
whence the name.

Using twistor techniques, Hitchin [\hita] has shown that ALE manifolds are
smooth resolutions of algebraic varieties in $\complex^3$. Simple singularities
admit an A-D-E classification according to which, the class of \mc of \GH
may be identified with the resolution of singularities of A-type.
A general construction of all ALE manifolds was worked out by
Kronheimer [\kro]. In the Kronheimer construction, these manifolds emerge
as minimal resolutions of $\complex^3/\Gamma$, where the discrete
subgroups of $SU(2)$, $\Gamma$, entering into the quotient are
in one-to-one correspondence
with the extended Dynkin diagrams of simply-laced simple Lie algebras.

\vskip 0.5truecm

\leftline{\bf Instanton Construction and Hyperk\"aler Quotient}

\REF\hklr{N.J.~Hitchin, A.~Karlhede, U.~Lindstr\"om and
M.~Ro\v cek, Comm. Math. Phys. {\bf 108} (1987) 535.}
ALE manifolds are non-compact \hk manifolds, \ie~manifolds which admit
three closed K\"ahler forms
$\w^i_{\mn} = g_{\m\l}(J^i)^\l{}_\n$, where $i=1,2,3$,
$g_{\mn}$ is the metric on the manifold and $J^i$ are three covariantly
constant complex structures satisfying the quaternionic algebra.
The \hk metric on an ALE space may be explicitly constructed
following a standard procedure known as \hkq [\hklr].
This is a general method to build a hyperk\"ahler manifold, $X$,
starting from another one, $M$, admitting ``triholomorphic" isometries.
These isometries are generated by vector fields, $v$, satisfying
${\cal L}_v \w^i \equiv i_v d\w^i + d(i_v \w^i) = 0$,
where $i_v$ denotes contraction with the components of $v$.
Any vector field of this kind admits in fact three Killing potentials,
$\m^i_v$, which can be thought as the \hk generalization of the well-known
Hamiltonian potentials corresponding to conserved quantities in classical
mechanics. They can be obtained integrating the equations $d\m^i_v = i_v \w^i$,
whenever the $\w^i$ are closed forms.

Let $M$ be a \hkm of real dimension $d_M=4m$ and $H$ a compact group
of $d_H=h$ freely acting \th isometries generated by $v_a$, $a=1,\dots h$.
One may construct a submanifold, $P_{\z}$, of dimension
$d_P = d_M - 3 d_H = 4m - 3h$, by defining
$$
P_{\z} = \{p\in M : \m^i_a(p) = \z^i_a, \quad i=1,2,3, \quad a=1,\dots h\}
\eqn\pmanifold
$$
When $\z^i_a$ belongs to $\real^3 \times {\cal Z}^*$, with ${\cal Z}^*$
the dual of the center of ${\cal H}$ (the Lie algebra of $H$), the
hypersurface $P_{\z}$ is preserved by the action of $H$.
In fact $P_{\z}$ turns out to be a $H$-principal bundle over the \hkm
$X_{\z}=P_{\z}/H$ of dimension $d_X = d_P - d_H = d_M - 4d_H = 4(m-h)$.
The coset space $X_{\z}$ is precisely the hyperk\"ahler quotient.

More explicitely, in the \Kro construction of ALE instantons, one starts
with the flat \hk space $M=\{ Q\otimes End(R) \}_\G$, \ie~ the space of
$\G$-invariant ``doublets'' of self-adjoint endomorphisms of
the vector space of the regular representation $R$ of $\G$.
Denoting by $R_i$ the irreducible representations of $\G$ of dimension $n_i$,
we have $R=\oplus_i R_i\otimes{\bar R}_i$ and $d_R = \sum_i n_i^2 = |\G|$,
where the index $i$ is taken to run from $0$ to $r-1$, with $r$ equal to
the number of conjugacy classes in $\G$.
The two-dimensional representation, $Q$, of $\G$ defines
the natural embedding of $\G$ into $SU(2)$ and it is isomorphic to the space
of left-handed spinors, $Q\sim S^-$. Its conjugate is denoted by $S^+$.
In taking the \hkq one identifies $H$ with the group $\prod_i U(n_i)/U(1)$ of
freely acting \th isometries commuting with the left action of $\Gamma$ on $R$.

\REF\gn{T.~Gocho and H.~Nakajima, J. Math. Soc. Jap. {\bf 44} (1992) 43.}
To the $H$-principal bundle, $P_{\z}$, one may associate the so-called
``\tb" $\ct$, a vector
bundle whose typical fiber is the vector space of the regular representation.
As a result of the above construction,
the curvature of the natural connection on the principal bundle
$P_{\z}$ over $X_{\z}$ is \sd [\hklr,\gn].

\REF\aty{For a review see M.~Atiyah, {\it Geometry of \YM fields}, Lezioni
Fermiane, Accademia Nazionale dei Lincei e Scuola Normale Superiore, Pisa
(1979).}

The ADHM construction [\adhm,\aty]
that gives rise to all self-dual connections on $S^4$ can also be cast in the
language of the \mom and carried over to $\real^4$.
Starting from this observation \Kro and \Nak brought the ADHM and the \hkq
construction of ALE spaces together [\kn].

The initial step in the KN construction
is to give a set of ``ADHM data" $\{A,B,s,t,\xi\}$ where $\xi\in M$,
$A$ and $B$ are $\G$-equivariant endomorphisms of a $k$-dimensional complex
vector space, $V$, and $s, t$ is a pair of homomorphisms between $V$ and
an $n$-dimensional
complex vector space $W$. Both $V$ and $W$ are $\G$-modules,\ie~
they admit the decompositions
$$
V=\oplus_i V_i\otimes R_i\qquad\quad W=\oplus_i W_i\otimes R_i
\eqn\decomp
$$
where $V_i\sim\complex^{v_i}$, $W_i\sim\complex^{w_i}$.
Therefore $k = dimV = \sum_i n_i v_i$ and $n = dimW = \sum_i n_i w_i$.

Out of these data a matrix ${\cal D}$ is defined by
$$
{\cal D}=({\cal A}\otimes \unita - \unita \otimes \xi) \oplus \Psi
\otimes \unita
\eqn\defd
$$
where
$$
{\cal A}=\pmatrix{A &-B^\dagger\cr B &A^\dagger\cr}\qquad\Psi=(s\quad
t^\dagger)
\qquad \xi=\pmatrix{\alpha &-\beta^\dagger\cr\beta &\alpha^\dagger}
\eqn\defa
$$
The $(2k+n)|\G|\times 2k|\G|$ matrix $\cd$ represents the linear map
$$
\cd:(S^+\otimes V\otimes\ct)\mapsto(Q\otimes V\otimes\ct)\oplus(W\otimes\ct)
\eqn\linmap
$$
To proceed in the KN construction of instantons on $X_{\z}$,
one should remember that $X_{\z}$ is a smooth resolution of  $\real^4/\G$.
One is thus lead to consider only the $\G$-invariant part of \linmap,
$\cd_{\G}$.
When $\cd$ is restricted in this way, it becomes a $(2k+n)\times 2k$ matrix
which is the analogue for ALE manifolds of the linear map, $D=a-bx$,
appearing in the ADHM construction on $\real^4$ [\adhm, \aty].

Self-duality of the resulting YM connection is imposed
by requiring the validity of the ADHM equations
$$
\eqalign{
[A,B]+st=-\zeta_{\complex}\cr
{i\over 2}([A,A^\dagger]+[B,B^\dagger])+s^\dagger
s-tt^\dagger=-\zeta_{\real}\cr
[\alpha,\beta]=\zeta_{\complex}\cr
{i\over 2}([\alpha,\alpha^\dagger]+[\beta,\beta^\dagger])=\zeta_{\real}\cr}
\eqn\adhmeq
$$
which ensure that $\cd^\dagger_{\G}\cd_{\G}=\Delta\otimes\unita$ with
$\Delta$ a
real $k\times k$ matrix and $\unita$ the identity acting in $S^+$.
We remark that the last two equations in \adhmeq~are precisely the equations
defining the submanifold $P_{\z} \subset M$ in eq.\pmanifold.
The instanton bundle on $X_{\z}$ is then identified with
$
{\cal E}={\rm Ker}\cd^\dagger_{\G}\subset({\cal Q}\otimes{\cal V}
\otimes\ct)_{\G}\oplus({\cal W}\otimes\ct)_{\G}$,
where ${\cal Q}, {\cal V}, {\cal W}$ denote the trivial vector
bundles over $X_\zeta$ with fiber $Q, V, W$ respectively and the
subscript $\G$ means restriction to $\G$-invariant subspaces.
{\cal E} is a complex vector bundle of rank $n$.

The YM connection on {\cal E} is finally given by
$$
A_\mu=U^\dagger\nabla_\mu U
\eqn\connec
$$
where $U$ is a $(2k+n)\times n$ matrix of orthonormal sections of
${\rm Ker}\cd_{\G}^\dagger$, \ie~a matrix obeying $\cd_{\G}^\dagger U=0$ and
$U^\dagger U = \unita_{n\times n}$.
Since ${\cal Q}, {\cal V}, {\cal W}$ are flat bundles,
the covariant derivative on $({\cal Q}\otimes{\cal V}\otimes\ct)_{\G}$
is simply given by $\nabla_\mu=(\partial_\mu+A_\mu^{\ct})$, with $A_\mu^{\ct}$
the (self-dual) connection on the tautological bundle with values in ${\cal
H}$.

The first and second Chern classes of ${\cal E}$ are given by the formulae
$$
c_1({\cal E}) = \sum_{i\neq \scriptscriptstyle 0} u_i c_1({\cal T}_i) \qquad
c_2({\cal E}) = \sum_{i\neq \scriptscriptstyle 0} u_i
c_2({\cal T}_i) + {dimV\over |\G|}
\eqn\chern
$$
where the vector bundles ${\cal T}_i$ are defined by
the decomposition ${\cal T} = \oplus_i {\cal T}_i \otimes R_i$
similar to the decompositions in eq.\decomp. In terms of the dimensions,
$w_i$ and $v_i$, of the vector spaces $W_i$ and $V_i$, the integers $u_i$
are defined by $u_i=w_i-C_{ij}v_j$, with $i,j=0,1,\dots r-1$
and $C_{ij}$ is the extended Cartan matrix of the Lie algebra (of
type A-D-E) associated to the discrete group $\G$.

The complex dimension
of the framed moduli space, ${\cal M}_\zeta$, which turns out to be a \hkm
[\kn,\nak], is
$$
{\rm dim}{\cal M}_\zeta= \sum_i u_i(v_i+w_i)
\eqn\purepeggio
$$
\REF\cg{E.~Corrigan and P.~Goddard, Ann. Phys. {\bf 154} (1984) 253.}
Generalizing the inverse construction of Corrigan and Goddard [\cg], \Kro
\par\noindent and \Nak have also shown the uniqueness and completeness
of the above construction [\kn].

We now specialize the discussion to the simplest case of the
minimal $SU(2)$ instanton on the EH gravitational background in \ehmetric.
In this case $\G=\zet_2$, the flat \hkm $M$ is $\real^8$ and the
\hkq is taken with respect to the group $H=U(1)$, since the two irreducible
representations of $\zet_2$ are one-dimensional
($n_{\scriptscriptstyle 0} = n_1 = 1$).
The minimal $SU(2)$ instanton bundle,
${\cal E}(k=1, n=2)$, corresponds to the choice $w=(0,2), v=(0,1),
u=(2,0)$ and, according to \chern,
has topological numbers $c_1({\cal E})=0, c_2({\cal E})=\mezzo$.
In this case the matrices $A$ and $B$, defined in \defa,
are absent, while
$$
\alpha=\pmatrix{0&x_1\cr y_1&0\cr}\qquad \beta=\pmatrix{0&x_2\cr
y_2&0\cr}\qquad
\Psi=\pmatrix{\sigma_1&-\bar\tau_2\cr\sigma_2&\bar\tau_1\cr}
\eqn\questaserve
$$
where $x_1$, $x_2$, $y_1$, $y_2$ are complex coordinates on $\real^8$,
$\s_1$, $\s_2$, $\t_1$, $\t_2$ are complex
parameters and the bar indicates complex conjugation.

The form of $\cd^\dagger$, when restricted to $\G$-invariant subspaces,
reduces to
$$
\cd^\dagger_{\G} = \pmatrix{\xbar_1 &\xbar_2 &\bar\sigma_1&\bar\sigma_2\cr
-y_2&y_1&-\tau_2&\tau_1\cr}
\eqn\cdrestricted
$$

Putting $\zeta_{\real}=-a^2,\zeta_{\complex}=0$
\foot{For our purposes, we are allowed to
eliminate the two moduli represented by $\zeta_{\complex}$
through a global rotation which corresponds to
a (non analytical) change of coordinates on the EH manifold.} and
defining $X^2=|x_1|^2+|x_2|^2$, $\rho^2=|\sigma_1|^2+|\sigma_2|^2$, the
ADHM equations are solved by
$y_1=\lambda x_1, y_2=\lambda x_2, \tau_1=\mu\sigma_1, \tau_2=\mu\sigma_2$
with $\lambda^2=1+{a^2\over X^2}, \quad\mu^2=1-{a^2\over\rho^2}$.
In these equations $\lambda$ and $\m$ have been chosen to be real.
This can always be done by exploiting the $U(1)$ isometries of the
principal bundles $P_{\z}$ and ${\cal P}_{\z}$ over $X_{\z}$ and
${\cal M}_{\z}$, respectively.
An orthonormal basis for ${\rm Ker}\cd_{\G}^\dagger$ can be checked to be
$$
U={1\over X\rho\sqrt{X^2+\rho^2}}\pmatrix{\rho^2 x_1&-\mu\rho^2\bar x_2\cr
\rho^2 x_2&\mu\rho^2\bar x_1\cr -X^2\sigma_1&\lambda X^2\bar\sigma_2\cr
-X^2\sigma_2&-\lambda X^2\bar\sigma_1\cr}
\eqn\ugauge
$$
In this setting, $SU(2)$ gauge transformations correspond to $U \to UN$
with $N \in SU(2)$.
The last ingredient needed to compute the self-dual connection \connec~
is the abelian ($H=U(1)$) connection on the \tb $\ct$.
With a proper gauge choice, $A_\mu^{\ct}$ may be identified with
the monopole potential given in \ehmonopole.
Switching to the coordinates employed in \ehmetric~ and inserting \ugauge~ in
\connec, one explicitely gets
$$
A = A_\mu dx^\m =
i\pmatrix{f(r)\sigma_z&g(r)\sigma_-\cr g(r)\sigma_+&-f(r)\sigma_z\cr}
\eqn\aconnec
$$
where $\sigma_\pm=\sigma_x\pm i\sigma_y$ and
$$
f(r)={t^2 r^2+a^4\over r^2(r^2+t^2)}\qquad g(r)={\sqrt{t^4-a^4}\over r^2+t^2}
\eqn\fg
$$
with $t^2=2\rho^2-a^2$ and $r^2=2X^2+a^2$.

\REF\bcc{H.~Boutaleb-Joutei, A.~Chakrabarti, A.~Comtet, Phys. Rev. {\bf D 21}
(1980) 979; ibid. 2280; ibid. 2285.}
The connection \aconnec~ was previously found in [\bcc]
following a completely different procedure. In the limit $a\to 0$ \aconnec~
becomes a connection over $\real^4/\zet_2$ and coincides with the BPST
instanton (in the singular gauge) with center at $x_0=0$ and size $t$ [\aty].

The requirement of $\zet_2$ invariance effectively reduces by a factor
of two, in agreement with \purepeggio, the dimension of the framed moduli
space ${\cal M}_\z$. It can be checked that ${\cal M}_\z$ exactly coincides
with the four-dimensional
EH manifold [\kn,\nak]. By supersymmetry we also conclude that the
number of spinor zero modes in the adjoint
representation of the $SU(2)$ gauge group is two,
in agreement with the index formulae given in [\nak].

The explicit expressions of the gaugino and gauge field zero modes
can be found starting from the three bounded harmonic scalars of isospin
$j=1$ in the minimal instanton background.
In the chosen gauge, which is equivalent to the so-called singular gauge,
these correspond to harmonics with angular momentum $l=0$.

Labelling the components of the scalar isovector by the
eigenvalues of $j_3$, they obey the equations
$$
\eqalign{
\Delta\theta_{j_3=\pm 1}&=\biggl[{1\over r^3}\ddr\biggl(r^3\bigl(
\unoar\bigr)\biggr)\ddr-(f^2+g^2)\biggr]\theta_{j_3=\pm 1}=0\cr
\Delta\theta_{j_3=0}&=\biggl[{1\over r^3}\ddr\biggl(r^3\bigl(
\unoar\bigr)\biggr)\ddr-2g^2\biggr]\theta_{j_3=0}=0\cr}
\eqn\scalarlapl
$$
whose bounded solutions are
$$
\theta_{j_3=0}={t^2r^2+a^4\over t^2(r^2+t^2)} \qquad
\theta_{j_3=\pm 1}={\sqrt{r^4-a^4}\over r^2+t^2}
\eqn\scalzero
$$

The spinor \zms in the adjoint representation of $SU(2)$ can now be
expressed in the form
$$
\l^j_{\a (\scriptscriptstyle 0\scriptstyle)}=
\s^\m_{\a\adot}(D_\m \theta)^j\bar\eps^{\adot}
\eqn\susygaugino
$$
where $\bar\eps^{\adot}$ is the covariantly constant spinor on the EH
background
and $\s^\m=(-i\unita, \s^j)$, with $\s^j$ the Pauli matrices.
Out of the six possible choices of $\theta^j$ and $\bar\eps^{\adot}$, only
two linearly independent zero-modes can be obtained. They can be written
in terms of a constant spinor $\eta_\a$ as
$$
\l^j_{\a (\scriptscriptstyle 0\scriptstyle)} = f^j\s^j_\a{}^\b \eta_\b
\eqn\funzioniuno
$$
where the index $j$ is not summed over and
$$
f^1=f^2={ 2(t^2r^2+a^4)\over r(r^2+t^2)^2}
\qquad
f^3={2\sqrt{(t^4-a^4)(r^4-a^4)}\over r(r^2+t^2)^2}
\eqn\funzioni
$$
The norm of the gaugino \zms defined in \funzioniuno~and \funzioni~is
computed to be ${\sqrt 2}\pi t$.

\vskip 0.5truecm

\leftline{\bf Instanton-dominated Correlation Functions}

\REF\hp{S.W.~Hawking, C.N.~Pope, Nucl. Phys. {\bf B146} (1978) 381.}
\REF\kmp{K.~Konishi, N.~Magnoli and H.~Panagopoulos,
Nucl. Phys. {\bf B309} (1988) 201; ibid. {\bf B323} (1989) 441.}
We now describe the computation of instanton dominated correlation functions
in SYM theory.
We recall that the functional integral is performed by expanding the SYM action
around its relevant minima up to quadratic terms in the field fluctuations.
The quadratic integration gives rise to the determinants of the bosonic and
fermionic kinetic operators  which, due to the global supersymmetry,
cancel each other up to \zms [\hp,\kmp].

According to the number of supersymmetries, we find the following results.

\vskip 0.5truecm

\leftline{\bf N=1}

The $N=1$ vector multiplet contains
a gauge field $A_\mu^i$ and a gaugino $\lambda_\alpha^i$
both in the adjoint representation of the $SU(2)$ gauge group.

The correlation function, which satisfies
the chiral selection rule associated to the anomaly of the R-symmetry
current in the minimal instanton saddle-point, is
$<{g^2\over 16\pi^2} {\rm Tr} \lambda\lambda(x)>$, with $g^2$ the square
of the gauge coupling constant inserted to get a renormalization
group invariant answer [\susy].

The two gaugino insertions are saturated by the two fermionic
zero modes \funzioniuno~
and the integration over the bosonic zero-modes is traded
for an integration over the four moduli of the minimal instanton.
The jacobian for this change of integration variables is
$$
\prod_{I=0}^3 {|| \delta_I A|| \over {\sqrt {2\pi}}}=
{64 \pi^4 t^3 \over ({\sqrt {2\pi}} g)^4}
\eqn\bosjacobi
$$
where the indices $0,1,2,3$ are associated to the collective coordinates
$t,\theta^1,\theta^2,\theta^3$, respectively. In detail one has
$||\delta_0 A||^2={8\pi^2 \over g^2}{t^4 \over t^4-a^4}$,
$||\delta_1 A||^2=||\delta_2 A||^2={8\pi^2 \over g^2} t^2$ and
$||\delta_3 A||^2={8\pi^2 \over g^2}{t^4-a^4 \over t^2}$.
After performing the integration over the fermionic partners, $\eta$, of
the moduli, which amounts to sum over the permutations of
the fermionic zero-modes, the gaugino condensate is computed to be
$$
\eqalign{
&<{g^2\over 16\pi^2} {\rm Tr}\lambda\lambda(x)>
 = e^{-{8\pi^2\over g^2}\mezzo}\mu^3\int_a^\infty
dt\int_{SU(2)/\zet_2} d^3\theta {64 \pi^4 t^3 \over ({\sqrt {2\pi}} g)^4}
\times\cr
&({{\sqrt 2}\over \pi t})^2 {g^2\over 32\pi^2}
\biggl[{2(t^2 r^2+a^4)^2\over r^2(r^2+t^2)^4}+
{(t^4-a^4)(r^4-a^4)\over r^2(r^2+t^2)^4}\biggr] = \mezzo \Lambda^3_{N=1}\cr}
\eqn\condensate
$$
where $\Lambda_{N=1}$ is the (2-loop) renormalization group invariant scale of
the $N=1$ SYM theory.

\vskip 0.5truecm

\leftline{\bf N=2}

\REF\west{For a review of extended supersymmetry see \eg~P.~West'
{\it Introduction to Supersymmetry and Supergravity}, World Scientific,
Singapore 1990.}
The field content of the $N=2$ vector multiplet is given by a gauge field
$A_\mu^i$, a doublet of gauginos $\lambda_\alpha^{Ai}$ and a complex scalar
$\phi^i$, all in the adjoint representation of the gauge group. Apart from
the non-anomalous $SU(2)$ global symmetry which rotate the two gauginos,
this theory admits an anomalous $U(1)$ R-symmetry under which the $\l^A$'s
have charge $+1$ and $\phi$ has charge $+2$ [\west].

In this case we have four gaugino zero modes and the correlation function
which is dominated by the \min is $<{\rm Tr}\phi^2(x)>$. In fact the lowest
order non-trivial contribution is found by
expanding the action to second order in the gauge coupling. In this
way two powers of the Yukawa interaction term (see eq.(25) below)
are brought down from the exponential: each scalar field insertion thus
counts as two gaugino zero modes [\susy]. The next step is to
perform the Wick contractions of the scalar fields
which effectively amounts to
substitute each $\phi^i$ with $\phi^i_{(\scriptscriptstyle 0\scriptstyle)}$,
the solution of the differential
equation
$$
\Delta\phi^i_{(\scriptscriptstyle 0\scriptstyle)} =
{g\over {\sqrt 2}} \varepsilon^i_{jk}\varepsilon_{AB}
\lambda^{Aj}_{\alpha (\scriptscriptstyle 0\scriptstyle)}
\lambda^{\alpha B k}_{(\scriptscriptstyle 0\scriptstyle)}
\eqn\laplacell
$$
where $\varepsilon$ is the antisymmetric Levi-Civita tensor.
In terms of the functions
$$
h^1=h^2 ={\sqrt{(t^4-a^4)(r^4-a^4)}\over 4(r^2+t^2)^2}
\quad h^3 = - {(a^4+t^4)r^2+2a^4t^2\over 4t^2(r^2+t^2)^2}
\eqn\gunzioni
$$
the components of $\phi^i_{(\scriptscriptstyle 0\scriptstyle)}$ are
$$
\phi^i_ {(\scriptscriptstyle 0\scriptstyle)} = -ig{\sqrt 2}
h^i\eta^\alpha_A (\sigma^i)_\alpha{}^\beta\eta^A_\beta
\eqn\chepalle
$$
where the index $i$ is not summed over and the $\eta^\alpha_A$'s are
constant spinors. As before, after performing the integration
over the fermionic variables, $\eta^\alpha_A$, which implements the sum over
the permutations of the gaugino zero-modes, the correlation function becomes
$$
\eqalign{
&<{\rm Tr}\phi^2(x)>
 = e^{-{8\pi^2\over g^2}\mezzo}\mu^2\int_a^\infty dt
\int_{SU(2)/\zet_2} d^3\theta {64 \pi^4 t^3 \over ({\sqrt {2\pi}} g)^4}
\times\cr
&({{\sqrt 2} \over \pi t})^4 ({{\sqrt 2}i g\over 4})^2
\biggl[ {((a^4+t^4)r^2+2a^4t^2)^2\over t^4(r^2+t^2)^4}
+{2(t^4-a^4)(r^4-a^4)\over (r^2+t^2)^4} \biggr] = - 2 \Lambda^2_{N=2}\cr}
\eqn\coorndue
$$
where $\Lambda_{N=2}$ is the renormalization group invariant scale of
the $N=2$ SYM theory.

\vskip 0.5truecm

\leftline{\bf N=4}

The $N=4$ supermultiplet contains the gauge vector boson $A_\mu^i$,
four gauginos $\lambda^A$ (in the fundamental representation, {\bf 4}, of
the $SU(4)$ group which rotates the four SUSY charges)
and six scalars $\phi_{AB}^i$ (in the antisymmetric
representation, {\bf 6}, of the $SU(4)$ group).
All the fields are in the adjoint representation of
the $SU(2)$ gauge group and the six scalars satisfy the reality condition:
$(\phi_{AB})^* = \phi^{AB} = \mezzo\varepsilon^{ABCD} \phi_{CD}$.
The would-be R-symmetry, which in $N=1$ notation corresponds to the abelian
factor in the manifest $SU(3)\otimes U(1)$ global symmetry, becomes part of
the global $SU(4)$ and is not anomalous [\west]. In fact, under this conserved
$U(1)$ current the gauginos in the three $N=1$ chiral multiplets have charge
$+1$, while the gaugino in the vector multiplet has charge $-3$.

\REF\sei{N.~Seiberg, Phys. Lett. {\bf 206B} (1988) 75.}
For this reason, the simplest correlation function which can receive
contribution
from the \min saddle-point is the vacuum expectation
value of the identity operator. For SYM theory in flat space, non-perturbative
corrections to the vacuum amplitude are expected to be zero [\sei]
and our results point in this direction (see below). Instead, in curved
backgrounds, the vacuum amplitude
is conjectured to be the sum of the Euler characteristics of the
relevant moduli spaces weighted with the exponential of the corresponding
instanton action [\vw].

The lowest non-trivial order in $g$ at which
it is possible to saturate the fermionic zero-modes is $g^4$ and the
functional integral effectively becomes
$$
<\unita>={g^4\over ({\sqrt 2})^4 4!}<\bigl(\int d^4x
\varepsilon^{ijk}\phi^i_{AB} \lambda^{jA} \lambda^{kB}\bigr)^4>
\eqn\coornquattro
$$
Performing the Wick contractions between pairs of scalar fields one is left
with the scalar propagator acting on the gaugino bilinears.
Integrating over two of the four positions amounts to substituting
$\phi$ with the solution \chepalle~of equation \laplacell.
After some algebraic manipulations, each one of the remaining two integrands
takes the form
$$
\sum_i h_i f_{i+1} f_{i+2}
\vareps_{ABCD}\eta^A\sigma^i\eta^B\eta^C\sigma^i\eta^D =
(h_3f_1f_2-h_1f_2f_3)
\varepsilon_{ABCD}\eta^A\sigma^3\eta^B\eta^C\sigma^3\eta^D
\eqn\manipolo
$$
We immediately notice that for $\real^4$ this quantity is identically
zero, because in this case $f_1=f_2=f_3$ and $h_1=h_2=h_3$, and
we conclude that the vacuum amplitude on $\real^4$ receives no-correction
from the minimal instanton sector.
In the EH background one gets instead a non-vanishing result.
Integrating \manipolo~ gives a result proportional to
$$
\int d^4x\sqrt{det(g^{EH})}(h_3f_1f_2-h_1f_2f_3) = - {\pi^2 a^4\over 8t^2}
\eqn\integrali
$$
where $det(g^{EH})$ stands for the determinant of the EH metric \ehmetric.
The integration over the fermionic collective coordinates,
$\eta_A$, simply gives a factor $4!$.
Performing the integral over the moduli one finally gets
$$
<\unita> =
e^{-{8\pi^2\over g^2}\mezzo}\int_a^\infty dt
\int_{SU(2)/\zet_2}d^3\theta {64 \pi^4 t^3 \over ({\sqrt {2\pi}} g)^4}
({{\sqrt 2} \over \pi t})^8 ({\pi^2 a^4\over 8t^2})^2
({ 4! 3 g^4\over 4!}) = {3\over2} e^{-{8\pi^2\over g^2}\mezzo}
\eqn\identita
$$
Recalling that ${\cal M}_{\z}$ coincides with the EH manifold,
we see that this result is consistent
with the value of the Euler characteristic of ${\cal M}_{\z}$,
if one identifies the above result with the ``bulk'' contribution to
$\chi({\cal M}_{\z})$ which is exactly $3/2$.

\vskip 0.5truecm

\leftline{\bf Conclusions}

In this letter we have reported some preliminary computations of instanton
dominated correlation functions in SYM theories on the EH space.
We have only studied the minimal instanton case
(second Chern class $c_2({\cal E})=\mezzo$) leaving the cases of higher
Chern class to a forthcoming paper in which the physical and the mathematical
aspects of our computations will be discussed at length [\bfmrb].
{}From the computations presented
here we can already draw some lessons. The constancy of instanton dominated
correlation functions of lowest components of chiral superfields seems to
persist even on ALE backgrounds. This fact is to be related to the existence
of a global supersymmetry charge associated to a covariantly constant Killing
spinor.

The independence of gaugino condensates from any scale (\eg~ $a$) other than
the SYM renormalization group invariant scale
bears immediate consequences on the supergravity
extension of such computations. A factorization between the gravitational
and gauge sectors can in fact be envisaged, making this extension trivial,
through a patching of the above computation with those of [\kmp].

The final goal of these considerations is to extract non-perturbative
corrections to the low-energy effective field theory emerging from
superstring. However, as we remarked before, the \min is a solution
of the superstring classical equation of motion which
requires to be adjusted order by order in $\ap$. From this point of view it
appears to be more interesting to consider the case in which no $\ap$
corrections to the lowest order classical solution are expected, as it happens
with the $k=3$ ($c_2={3\over 2}$) instanton, thanks to the standard embedding.

In the $N=2$ case, the calculation performed above should give the first
instanton correction to the analytic prepotential of a SYM theory
on the EH space. In principle it should be possible to
sum up the contributions of gauge instantons with higher Chern class
on ALE spaces, in analogy to what was done in [\sw].

We would like to stress that all computations presented here were performed
with
zero expectation value of the scalar fields (we do not
expect the costancy of the correlators to be influenced by this choice, unlike,
perhaps, the contants in front of the r.h.s of eqs.
\condensate, \coorndue, \identita). The case in which the scalar
{\it vev}'s are different from zero deserve further study.

A last remark concerns the $N=4$ theory, for which we would like to draw a
comparison with the results of [\ggpz]. Since we have imposed the same
boundary conditions on bosonic and fermionic fields in order to
preserve SUSY, our results should be considered as corrections to the
Witten index rather than to the free energy.

\endpage
\refout
\end

\REF\fh{M.~Furuta, Y.~Hashimoto, J. Fac. Sci. Univ. Tokyo {\bf 37} (1990) 585.}
\REF\hitb{N.~Hitchin, {\it Metrics on Moduli Spaces}, in Proceedings of the
Lefschez Centennial Conference (Contemporary Math. 58 Part I),
A.M.S. Providence R.I. 1986.}
\REF\fu{D.S.~Freed and K.K.~Uhlenbeck, {\it Instantons and Four-Manifolds},
MSRI Publ. 1, Springer-Verlag, 1984.}
\REF\egh{T.~Eguchi, P.B.~Gilkey and A.J.~Hanson, Phys.Rep. {\bf 66} (1980)
213.}
\REF\wsusy{E.~Witten, Nucl.Phys. {\bf B185} (1981) 513.}
\REF\aftop{D.~Anselmi and P.~Fre', Nucl.Phys. {\bf B392} (1993) 401;
ibid. {\bf B404} (1993) 288; ibid. {\bf B416} (1994) 255.}
\REF\sym{V.~Novikov, M.~Shifman, A.~Vainshtein and
V.~Zakharov, Nucl.Phys. {\bf B260} (1985) 157; \nextline
D.~Amati, G.C.~Rossi and G.~Veneziano, Nucl.Phys. {\bf B249} (1985) 1;\nextline
I.~Affleck, M.~Dine and N.~Seiberg, Nucl.Phys. {\bf B256} (1985) 557.}
\REF\agnt{I.~Antoniadis, E.~Gava, K.~Narain, T.~Taylor,
{\it Topological Amplitudes in String Theory}, preprint IC/93/202.}